# Bond-Selective Full-Field Optical Coherence Tomography

HAONAN ZONG[1], CELALETTIN YURDAKUL[1], JIAN ZHAO[1,4], ZIAN WANG[2], FUKAI CHEN[3], M. SELIM ÜNLÜ[1,2], AND JI-XIN CHENG[1,2,*]

[1]*Department of Electrical and Computer Engineering, Boston University, Boston, MA 02215, USA*
[2]*Department of Biomedical Engineering, Boston University, Boston, MA 02215, USA*
[3]*Department of Biology, Boston University, Boston, MA 02215, USA*
[4]*Current affiliation: The Picower Institute for Learning and Memory, Massachusetts Institute of Technology, Cambridge, Massachusetts 02142, USA*
*\* jxcheng@bu.edu*

**Abstract:** Optical coherence tomography (OCT) is a label-free, non-invasive 3D imaging tool widely used in both biological research and clinical diagnosis. Current OCT modalities can only visualize specimen tomography without chemical information. Here, we report a bond-selective full-field OCT (BS-FF-OCT), in which a pulsed mid-infrared laser is used to modulate the OCT signal through the photothermal effect, achieving label-free bond-selective 3D sectioned imaging of highly scattering samples. We first demonstrate BS-FF-OCT imaging of 1 μm PMMA beads embedded in agarose gel. Next, we then show 3D hyperspectral imaging of polypropylene fiber mattress from a standard surgical mask. We then demonstrate BS-FF-OCT imaging on biological samples, including cancer cell spheroids and *C. elegans*. Using an alternative pulse timing configuration, we finally demonstrate the capability of BS-FF-OCT on a bulky and highly scattering 150 μm thick mouse brain slice.

## 1. Introduction

Since the first report by Huang *et al.* in 1991, optical coherence tomography (OCT) has experienced many advanced technical developments and demonstrated significant applications in the past decades. [1] OCT has evolved from time-domain OCT (TD-OCT) [2], which mechanically scans the optical phase of the reference arm to obtain the signal from different depths, to spectral-domain/Fourier-domain OCT (SD/FD-OCT) [3-5], which spectrally resolves the detected interferometric signal from different depths without mechanically scanning. SD/FD-OCT has dramatically improved the sensitivity and imaging speed of OCT and achieved in vivo retinal imaging [4] till video rate [5]. However, neither TD-OCT nor SD/FD-OCT is suitable for obtaining high-resolution en-face images of samples because these modalities acquire the signal from different depths at a fixed lateral location first and then scan the sample laterally. To enable high-resolution en-face OCT imaging, time-domain full-field OCT (FF-OCT) was developed. [6, 7] FF-OCT adopts widefield illumination and a multi-pixel detector (a CCD or CMOS camera) to obtain en-face images at a given depth without scanning across the sample. FF-OCT was applied to in vivo human corneal [8] and retinal imaging [9] for ophthalmic diagnosis. FF-OCT was also used for histological imaging of different types of tissues, such as human skin tissue [10], breast tissue [11], and brain tissue [12], for cancer diagnosis. However, those conventional FF-OCT modalities can only provide tomography images without any molecular information, which limits their potential applications to samples that have different chemical compositions but similar morphology.

Vibrational microscopy has been a widely used tool for label-free molecular imaging without sample perturbation. [13] In these techniques, Raman scattering, or linear infrared absorption, is measured to provide the contrast. More recently, the relatively weak signal and low acquisition speed of the spontaneous Raman scattering [14] have been boosted by coherence Raman scattering microscopy [15, 16]. Compared to Raman scattering, which has an extremely small cross-section ($\sim 10^{-30}$ to $10^{-28}$ cm$^2$), linear infrared (IR) absorption has ten orders of magnitude larger cross-section ($\sim 10^{-18}$ cm$^2$). Despite the large cross-section, conventional IR imaging technique such as Fourier transform infrared (FTIR) [17, 18] has poor spatial resolution due to the long illumination wavelength. To break this limitation, mid-infrared photothermal (MIP) microscopy, which indirectly measures the IR absorption by using the photothermal effect, was developed recently. [19, 20] Since then, MIP microscopy has evolved from point-scan [19-26] to widefield configurations [27-37]. As reviewed recently [38, 39], MIP microscopy offers a few advantages. First, sub-micron spatial resolution is achieved through the visible probe beam. Second, widefield MIP microscopy enables high-throughput chemical imaging by exploiting the advantage that linear IR absorption doesn't require a tight focus. By using widefield illumination and detection configuration, the imaging speed could reach half of the camera frame rate. Third, volumetric chemical imaging is possible through mid-infrared photothermal phase tomography [32, 33, 37]. Despite these advances, phase tomography, including optical diffraction tomography [32] and intensity diffraction tomography [37], is limited to weakly scattering samples and can't be applied to highly scattering specimens such as tissues.

Using OCT as the probe of the mid-infrared (MIR) photothermal effect can potentially enable bond-selective 3D imaging for highly scattering samples. Notably, both "photothermal" and "MIR" processes have been applied to OCT separately. On the one hand, photothermal OCT has been a powerful functional extension of OCT since its first demonstration by Fujimoto et al. in 2008. [40] Photothermal OCT is realized by adding another modulated heating beam to OCT and measuring the modulation of the OCT signal induced by the heating beam. Since the heating beam is also in the visible or near-infrared region, it provides limited molecular specificity to OCT by detecting signals from specific absorbers at the heating wavelength, which can be endogenous pigments [41-43] in the sample or exogenous contrast agents [40, 44-49] that are imported into the sample. Although the exogenous contrast agents can improve the molecular specificity, perturbations may be introduced to the sample during the labeling process. On the other hand, OCT in the "MIR" domain has been reported, including conventional OCT modalities using MIR light sources to improve penetration depth, [50] or a time-gated method to detect the reflection of MIR light from different depths, [51] while these techniques still suffer from the intrinsic MIR resolution limitation, which is the same as FTIR. Despite these efforts, bond-selective OCT that harnesses the MIR photothermal effect has not been reported.

In this work, we report bond-selective full-field optical coherence tomography (BS-FF-OCT), in which a pulsed MIR laser modulates the full-field OCT signal through the photothermal effect. Our technique enables label-free bond-selective 3D sectioning imaging of highly scattering thick samples. To achieve this, we integrate a modulated MIR heating beam into a time-domain FF-OCT. We use a broadband light-emitting-diode (LED) as the probe light source and a virtual lock-in camera as the detector [27]. Our system can measure the change in the OCT signal as a result of thermal expansion and refractive index change induced by MIR heating. First, we demonstrate 3D bond-selective imaging of 1 μm PMMA beads embedded in agarose gel, which confirms the isotropic 1-micron resolution of BS-FF-OCT. Second, we show 3D hyperspectral imaging of a polypropylene fiber mattress from a standard surgical mask and the comparison between BS-FF-OCT and FTIR to confirm the spectrum fidelity. Then, we demonstrate bond-selective volumetric imaging on biological samples, including cancer cell spheroids and *C. elegans*. Finally, we demonstrate the capability of the BS-FF-OCT setup on a very bulky and highly scattering biological sample, i.e., a 150-μm thick mouse brain tissue slice, using an alternative pulse timing configuration.

## 2. Results and discussion

*2.1 BS-FF-OCT principles, instrumentation, and image reconstruction*

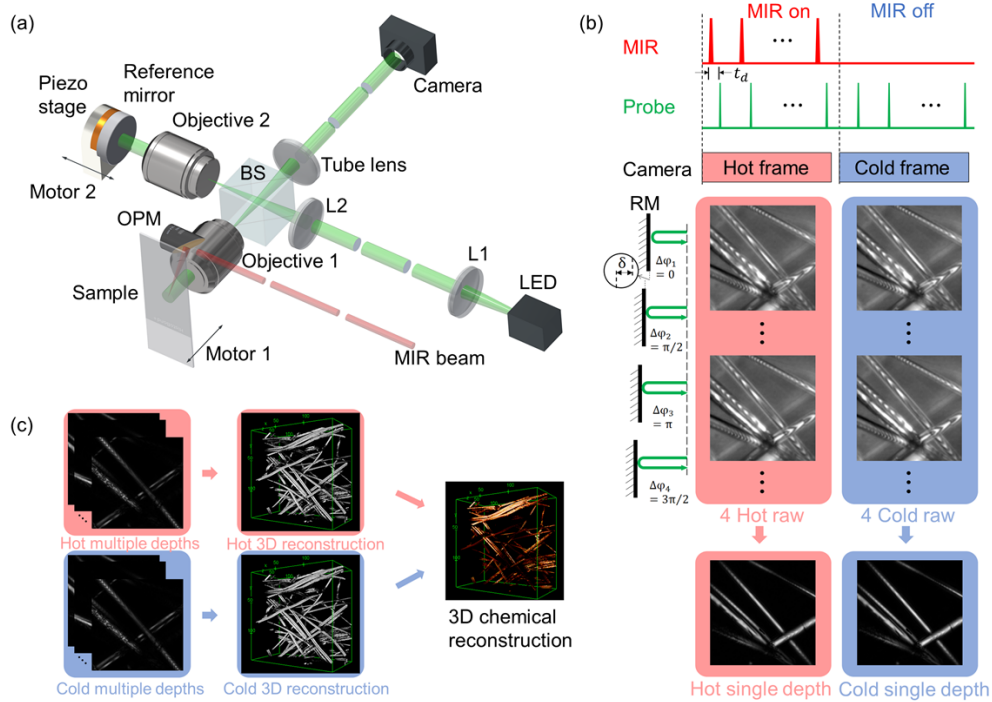

**Figure 1. BS-FF-OCT setup, synchronization, and image processing. (a) BS-FF-OCT setup configuration.** BS: Beam-splitter. L1-2: Lens. LED: Light-emitting-diode. OPM: Off-axis parabolic mirror (90 degrees). **(b) Synchronization and image acquisition at a single depth.** RM: reference mirror. Camera captures "hot" and "cold" frames, where the MIR beam is respectively on and off in a sequence. MIR and probe pulses are synchronized, and the time delay ($t_d$) between them is optimized to detect the maximum photothermal signal. Reference mirror is shifted a certain distance (δ) 4 times to create 4 pairs (hot and cold) interference raw images, and then a pair of FF-OCT images at a specific depth is obtained by image processing. **(c) Workflow of 3D image reconstruction.** By combining FF-OCT images at multiple depths, 3D reconstruction images can be obtained. Finally, 3D bond-selective image can be obtained by subtracting the hot and cold 3D images.

BS-FF-OCT relies on the modulation of the OCT signal by the photothermal effect induced by the MIR beam. The setup shown in **Fig. 1a** is compartmentalized into two sub-systems: (1) FF-OCT and (2) MIR modulation. For the FF-OCT part, the light source is a broadband light-emitting-diode (LED, central wavelength: 545 nm, FWHM: 100 nm). The reference mirror (reflectivity: 4%) is placed on a piezo scanner to create phase shifting between the reference and sample arms. Both the sample and reference mirrors are installed on motorized stages to scan different depths of the sample. For the MIR modulation part, a tunable MIR laser from 1320 cm$^{-1}$ to 1775 cm$^{-1}$ (linewidth: 10 cm$^{-1}$), covering the fingerprint region is used. The MIR and probe beams illuminate the sample from the same side.

The setup captures the depth-resolved photothermal FF-OCT images at a specific depth of the sample using a virtual lock-in technique [27], as shown in **Fig. 1b**. The top panel of **Fig. 1b** shows the timing configuration of the probe, MIR pulses, and camera exposure. The MIR pulse has a 20 kHz repetition rate and is modulated to "on" and "off" duty cycles by an optical chopper at 50 Hz. The probe pulse repetition rate is also set to 20 kHz which is synchronized with the MIR pulse with a specific delay time to optimize the photothermal signal. The camera frame rate is 100 Hz and is synchronized with the modulated "on" and "off" duty cycles of the MIR pulse. The camera-captured frames that correspond to the "on" and "off" duty cycles are called "hot" and "cold" frames, respectively. The middle panel of **Fig. 1b** shows that at each phase position of the reference mirror, a set of "hot" and "cold" raw frames are captured (to be averaged to 1 "hot" frame and 1 "cold" frame), and there are in total 4 phase positions. The bottom panel of **Fig. 1b** shows that 1 "hot" or "cold" FF-OCT image is obtained from the 4 "hot" or "cold" averaged raw frames, using the 4-frame phase-shifting algorithm [7]. Then, the depth-resolved photothermal FF-OCT image at this specific depth can be obtained by subtracting the "hot" and "cold" FF-OCT images. Furthermore, to obtain 3D reconstructed images for both hot and cold states, as shown in **Fig. 1c**, the sample is scanned at different depths with automatic coherence plane correction within the imaging volume (see details in the methods section). A 3D bond-selective OCT map can be obtained by subtracting the hot and cold 3D reconstructed images.

*2.2 BS-FF-OCT system characterization*

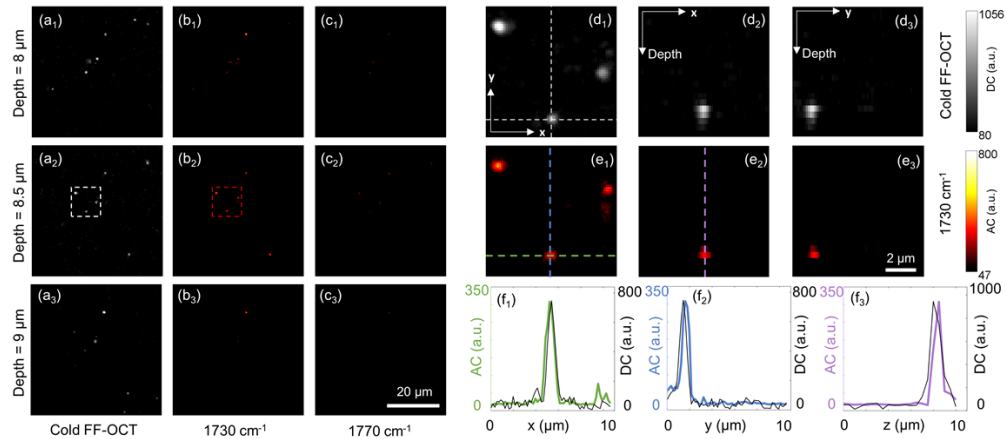

Figure 2. BS-FF-OCT imaging of 1 μm PMMA beads embedded in agarose gel. (a) cold FF-OCT images at different depths. (b-c) BS-FF-OCT images at 1730 cm$^{-1}$ and 1770 cm$^{-1}$. 1730 cm$^{-1}$ is the C=O band in PMMA, and 1770 cm$^{-1}$ is at off-resonance. (d$_1$) zoom-in view of the white dashed square area in (a$_2$). (d$_{2-3}$) cross-sectional images along the dashed lines in (d$_1$). (e$_{1-3}$) corresponding BS-FF-OCT images at 1730 cm$^{-1}$ of (d$_{1-3}$). (f$_{1-3}$) 1D cross-line profiles corresponding to the colored dashed lines in (e$_{1-2}$). The black profiles correspond to the cold FF-OCT images. FWHM of line profiles: 942.5 nm (green in (f$_1$)), 824.4 nm (black in (f$_1$)), 787.3 nm (blue in (f$_2$)), 772.7 nm (black in (f$_2$)), 870.3 nm (purple in (f$_3$)) and 1156.7 nm (black in (f$_3$)). BS-FF-OCT images are normalized by MIR powers.

To characterize the BS-FF-OCT setup, we first demonstrate 3D bond-selective imaging of 1 μm Poly(methyl methacrylate) (PMMA) beads embedded in agarose gel. **Fig. 2** shows that BS-FF-OCT achieves label-free volumetric vibrational spectroscopic imaging at isotropic 1-micron resolution. Specifically, **Fig. 2a-c** shows the cold FF-OCT, on-resonance, and off-resonance BS-FF-OCT images captured at three different depths with 0.5 μm step size. First, the cold FF-OCT images in **Fig. 2a** distinguish beads suspended at different depths (i.e., 1 μm apart), showing the depth-resolving capability of BS-FF-OCT setup. Second, to demonstrate the bond-selective capability, the MIR beam is set to an on-resonance absorption peak of PMMA at 1730 cm$^{-1}$. The BS-FF-OCT images show consistent features as in cold FF-OCT images (see **Fig. 2a-b**). Yet, the off-resonance BS-FF-OCT images at 1770 cm$^{-1}$ display no beads, as shown in **Fig. 2c**. **Fig. 2d-e** are the zoom-in views of a selected imaging 3D volume from three different directions. **Fig. 2d$_1$** and **Fig. 2e$_1$** are the corresponding areas indicated by the dashed squares in **Fig. 2a$_2$** and **Fig. 2b$_2$**, respectively. It can be seen from **Fig. 2d-e** that the beads have a slightly longer dimension along the optical axis. To characterize the axial and lateral resolution quantitatively, the 1D line profiles across the selected bead are plotted in **Fig. 2f**. The full-width half maximum (FWHM) of these line profiles are as follows, 942.5 nm (green in (f$_1$)), 824.4 nm (black in (f$_1$)), 787.3 nm (blue in (f$_2$)), 772.7 nm (black in (f$_2$)), 870.3 nm (purple in (f$_3$)) and 1156.7 nm (black in (f$_3$)). This result demonstrates the isotropic 1-μm resolution of the BS-FF-OCT setup. As a pump-probe technique, the resolution of the BS-FF-OCT setup is determined by the wavelength and optics of the probe beam [19]. For FF-OCT, the axial resolution ($\Delta z$) [7] can be calculated as $\Delta z = \left(\frac{1}{\Delta z_s^2} + \frac{1}{\Delta z_{NA}^2}\right)^{-\frac{1}{2}}$, where $\Delta z_s = \frac{2\ln(2)}{n\pi} \cdot \frac{\lambda_0^2}{\Delta \lambda}$ and $\Delta z_{NA} = \frac{n\lambda_0}{NA^2}$. The lateral resolution ($\Delta r$) can be calculated as $\Delta r = \frac{\lambda_0}{2 \cdot NA}$. Substituting $\lambda_0 = 545\ nm, \Delta\lambda = 100\ nm, n = 1, NA = 0.35$, the theoretical axial resolution $\Delta z$ can be calculated to be 1257.3 nm, and the theoretical lateral resolution $\Delta r$ can be calculated to be 778.6 nm. The theoretical axial and lateral resolution values are roughly consistent with the experimental FWHM values shown in **Fig. 2f**.

## 2.3 BS-FF-OCT imaging of polypropylene fiber mattress

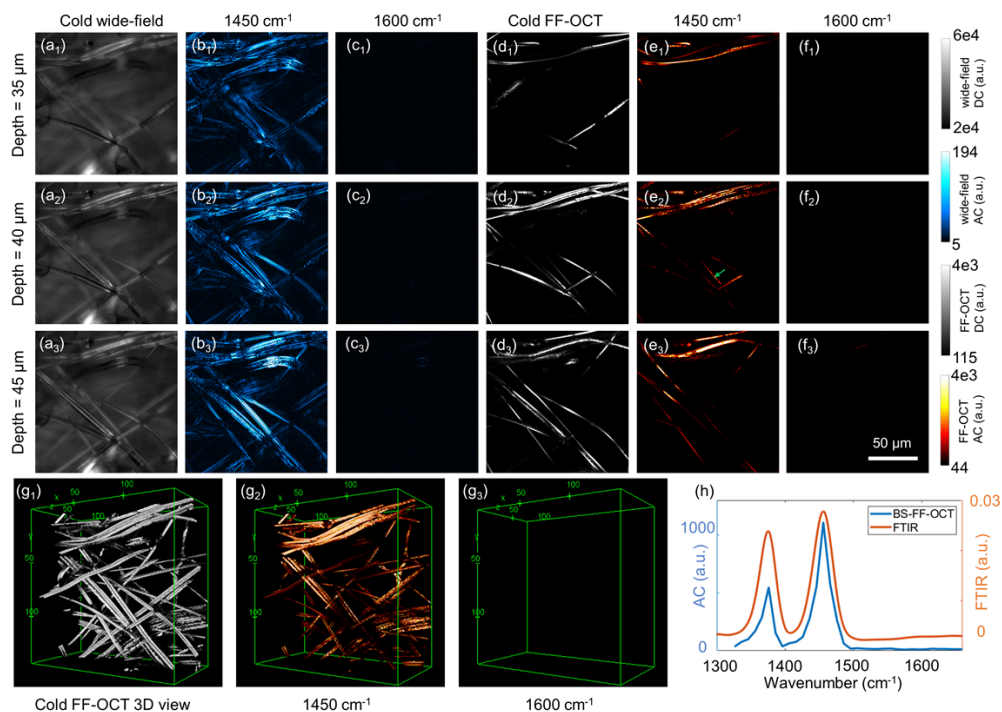

**Figure 3. BS-FF-OCT imaging of polypropylene fiber mattress. (a) cold widefield images at different depths. (b-c) widefield MIP images at 1450 cm$^{-1}$ and 1600 cm$^{-1}$.** 1450 cm$^{-1}$ is the C-H asymmetric deformation vibration bond in polypropylene, and 1600 cm$^{-1}$ is at off-resonance. **(d) cold FF-OCT images at different depths. (e-f) BS-FF-OCT images at 1450 cm$^{-1}$ and 1600 cm$^{-1}$. (g) 3D reconstruction of cold FF-OCT and BS-FF-OCT images. (h) comparison of BS-FF-OCT and FTIR spectrum.** The BS-FF-OCT spectrum is extracted from the position in (e$_2$) indicated by the green arrow. FTIR spectrum is acquired by a commercial FTIR spectroscopy from a bulky measurement of the polypropylene fiber sample. BS-FF-OCT images and spectrum are normalized by MIR powers. All images are denoised by BM4D algorithm. BS-FF-OCT and FTIR spectrum is smoothed by Gaussian-weighted moving average filter.

To demonstrate the 3D spectroscopic imaging capability of BS-FF-OCT, we use polypropylene fiber mattress from a standard surgical mask in air as a testbed (**Fig. 3**). To emphasize the depth-resolving capability of BS-FF-OCT, the cold, on-resonance, and off-resonance MIP images at different depths are captured as shown in **Fig. 3a-c**. Those widefield images are obtained under the same experimental condition and acquisition parameters except that the reference arm is blocked, which makes a fair comparison to those of BS-FF-OCT. As shown in **Fig. 3a-c**, the depth-resolving capability of conventional widefield MIP imaging is very limited, where the fiber features are indistinguishable. In contrast, BS-FF-OCT images in **Fig. 3d-f** clearly resolve features at different depths. Both widefield MIP images and BS-FF-OCT images demonstrate bond-selective capability, i.e., at the C-H asymmetric deformation vibration bond at around 1450 cm$^{-1}$. While **Fig. 3b** and **Fig. 3e** both show bright contrast, no contrast was found at the 1600 cm$^{-1}$ off-resonance wavenumber images (see **Fig. 3c** and **Fig. 3f**). To further show the 3D imaging capability of BS-FF-OCT, we perform 3D reconstruction of the polypropylene fiber mattress for a total depth range of 75 μm (see **Fig. 3g**). We notice that each fiber strip in **Fig. 3g** shows "double strips" which can be seen more clearly in the **Mov. S1a and Mov. S1b**. Since FF-OCT measures back reflections from the sample, the air-polypropylene top and polypropylene-air bottom interfaces of each fiber strip create two distinguishable strips. Also, the diameter of each fiber strip is larger than the axial resolution of the setup thus we can see the two reflection interfaces. **Fig. 3h** shows the BS-FF-OCT spectrum extracted from the position indicated by the green arrow in **Fig. 3e$_2$** and comparison with the FTIR spectrum. Both

BS-FF-OCT and FTIR spectra show peaks for the C-H symmetric deformation vibration bond at around 1370 cm$^{-1}$ and the C-H asymmetric deformation vibration bond at around 1450 cm$^{-1}$. These results further verify the bond-selective capability and demonstrate good spectral fidelity.

*2.4 BS-FF-OCT imaging of human bladder cancer cell spheroids and C. elegans*

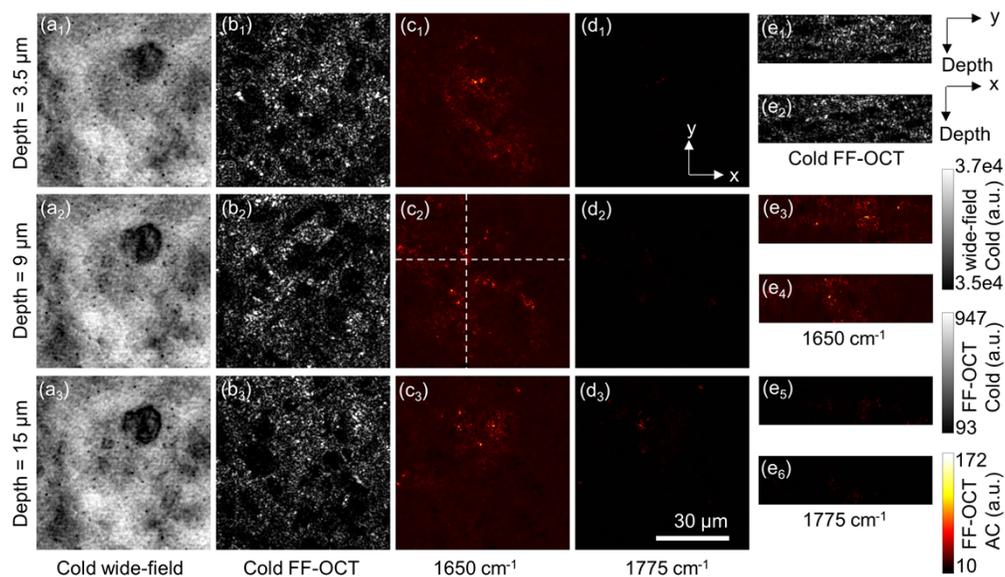

**Figure 4. BS-FF-OCT imaging of cancer cell spheroids. (a) Cold widefield images at different depths. (b) cold FF-OCT images at different depths. (c-d) BS-FF-OCT images at 1650 cm$^{-1}$, and 1775 cm$^{-1}$.** 1650 cm$^{-1}$ is the amide I band in protein, and 1775 cm$^{-1}$ is at off-resonance. **(e) Cross-sectional images along the dashed lines in (c$_2$).** BS-FF-OCT images are normalized by MIR powers.

To demonstrate the broad application potential of our technique on biological samples, we used human bladder cancer cell spheroids and *C. elegans* as testbeds. **Fig. 4** shows the BS-FF-OCT images of human bladder cell spheroids. The high-density areas (cytoplasm) and low-density areas (nucleus) inside the cell spheroids volume can be seen clearly (see **Fig. 4b**). Features from different depths can be distinguished compared to the cold widefield images in **Fig. 4a**. **Fig. 4c,** and **Fig. 4d** confirm the bond-selective capability, i.e., at 1650 cm$^{-1}$ (see **Fig. 4c**) in resonance with the amide I band of proteins where there is a stronger photothermal contrast than that of at off-resonance 1775 cm$^{-1}$ (see **Fig. 4d**). Moreover, the cutting-through sectioning images along the axial direction of the dashed lines in **Fig. 4c$_2$** show the cytoplasm and nucleus areas from the side views (**see Fig. 4e**).

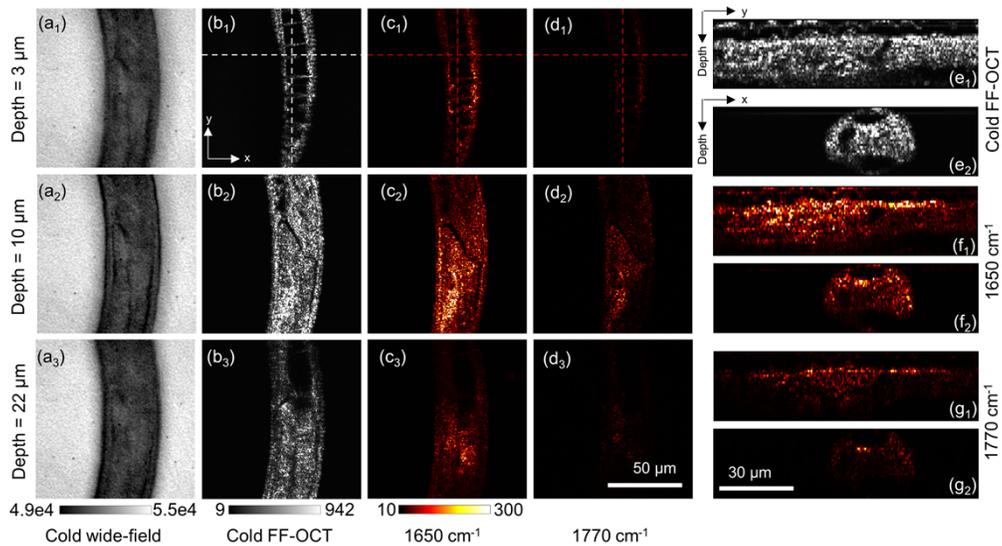

**Figure 5. BS-FF-OCT imaging of *C. elegans*. (a) cold widefield images at different depths. (b) cold FF-OCT images at different depths. (c-d) BS-FF-OCT images at 1650 cm$^{-1}$, and 1770 cm$^{-1}$.** 1650 cm$^{-1}$ is the amide I band in protein, and 1770 cm$^{-1}$ is at off-resonance. **(e-g) cross-sectional images along the dashed lines in ($b_1$), ($c_1$), and ($d_1$).** BS-FF-OCT images are normalized by MIR powers.

**Fig. 5** shows the BS-FF-OCT images of *C. elegans*. The cold FF-OCT images in **Fig. 5b** show features inside the *C. elegans* worm at various depths. In contrast, scatterers from different planes hinder these futures in the cold widefield images due to the lack of optical-sectioning capability (see **Fig. 5a**). The BS-FF-OCT images in **Fig. 5c** show strong photothermal contrast at 1650 cm$^{-1}$, amide I band whereas the photothermal contrast at the 1770 cm$^{-1}$ off-resonance wavenumber in **Fig. 5d** is weak. This confirms the chemical selective capability since the *C. elegans* is rich in protein. To further demonstrate the 3D sectioning capability of the BS-FF-OCT setup, the cutting-through sectioning images along the axial direction and dashed lines shown in **Fig. 5$b_1$-$d_1$** are plotted in **Fig. 5e-g**. In these side views, the different structures inside the worm are shown more clearly.

*2.5 BS-FF-OCT imaging of myelinated axons in mouse brain tissue*

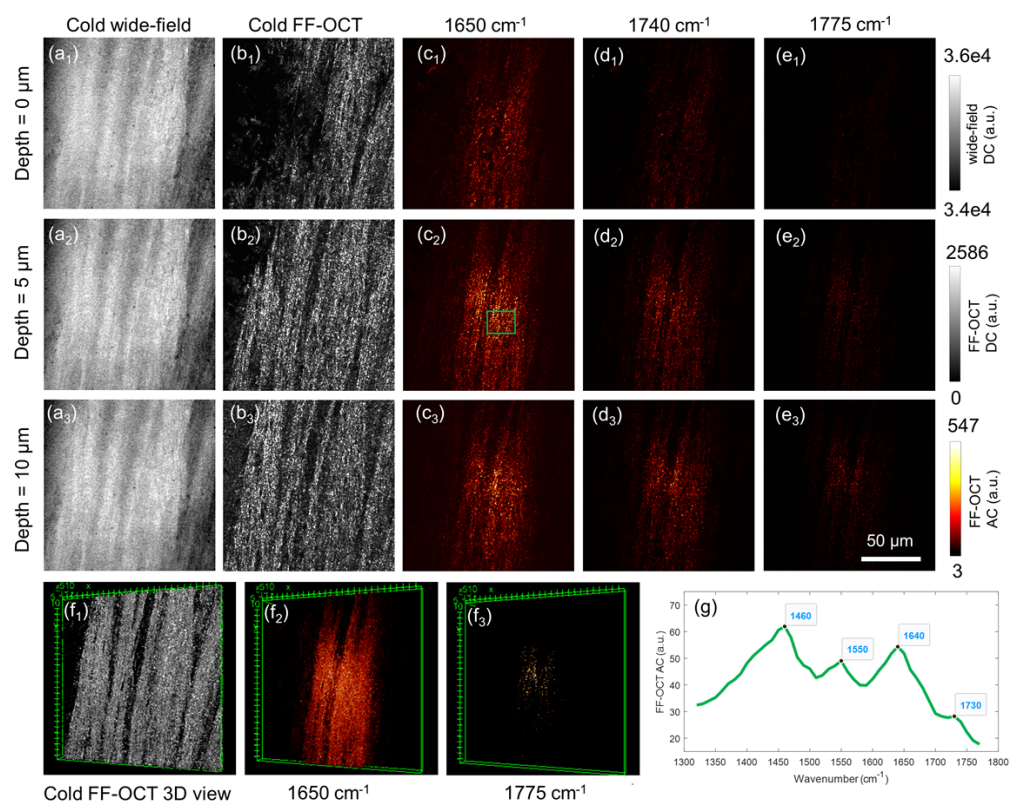

**Figure 6. BS-FF-OCT imaging of myelinated axons in mouse brain tissue. (a) cold widefield images at different depths. (b) cold FF-OCT images at different depths. (c-e) BS-FF-OCT images at 1650 cm$^{-1}$, 1740 cm$^{-1}$, and 1775 cm$^{-1}$.** 1650 cm$^{-1}$ is the amide I band in protein, 1740 cm$^{-1}$ is the C=O band in lipids, and 1775 cm$^{-1}$ is at off-resonance. **(f) 3D reconstruction of cold FF-OCT and BS-FF-OCT images. (g) BS-FF-OCT spectrum.** The BS-FF-OCT spectrum is extracted from the area in (c$_2$) indicated by the green rectangle. BS-FF-OCT Images and spectrum are normalized by MIR powers. All images are denoised by BM4D algorithm. BS-FF-OCT spectrum is smoothed by Gaussian-weighted moving average filter.

We chose a 150-µm thick mouse brain slice (**Fig. 6**) as the testbed to demonstrate the application potential of our setup for imaging thick and highly scattering biological samples. Martin Schnell *et al.* previously demonstrated infrared spectroscopic imaging of biological tissues through a Mirau interference objective. [31] The tissues used in that work were 5-µm-thin paraffin-embedded-sliced tissues prepared through a complex protocol. In this work, we demonstrate BS-FF-OCT imaging of fresh 150-µm thick brain slices sectioned by a simple procedure. The BS-FF-OCT setup can image thicker tissues owing to its particular design. The reference arm is fixed in the study by Martin Schnell *et al.* [31], since a Mirau objective is adopted to generate the interference signal. In contrast, BS-FF-OCT is based on a time-domain FF-OCT with a separated and tunable reference arm. Thus, in our BS-FF-OCT setup, the coherence plane can be tuned to the deeper layers of the samples. **Fig. 6a** shows the cold widefield reflection images focused at different depths. Due to limited depth-resolving capability, **Fig. 6a** looks similar at all depths. The photothermal widefield reflection images focused at different depths shown in **Fig. S1** also look similar. In comparison, the cold FF-OCT brain tissue images can distinguish myelinated axon structures from different depths (see **Fig. 6b**).

Second, an alternative MIR and probe pulse timing configuration is adopted to maximize the detected photothermal signal. Simulations by Zong *et al.* demonstrate that photothermal cooling time increases with the sample size [36]. The maximum photothermal signal can be obtained when the temperature difference between the "hot" and "cold" states is largest. Therefore, there should be enough time between the probe pulses to differentiate the "hot" and "cold" states. In the pulse timing configuration shown in **Fig. 1b**, the time window between the first probe pulse for the "cold" state and the last probe pulse for the "hot" state is only 50 μs, which is not enough for the cooling of the 150-μm-thick brain tissue. Thus, a new timing configuration is added to the setup, as shown in **Fig. S3**. The maximum cooling time in this alternative timing configuration is limited to 10 ms by the camera period time. MIR-probe delay scan is also performed, as shown in **Fig. S4** and **Fig. S5**. The cooling time constant of the 150-μm-thick brain tissue is found to be about 1.21 ms, showing that this thick tissue sample indeed requires an alternative timing configuration. Using the optimized MIR-probe delay value shown in **Fig. S5a**, BS-FF-OCT imaging results of myelinated axons of different depths are shown in **Fig. 6c-e**. At the 1650 $cm^{-1}$ Amide I and 1740 $cm^{-1}$ C=O bands, the BS-FF-OCT contrast is strong whereas the images at the 1775 $cm^{-1}$ off-resonance wavenumber have very weak contrast. This result reflects the major chemical content of myelinated axons, i.e., protein and lipids. The 3D reconstruction results of the cold FF-OCT and BS-FF-OCT images at 1650 $cm^{-1}$ and 1775 $cm^{-1}$ are shown in **Fig. 6f**.

To demonstrate the chemical selectivity, hyperspectral BS-FF-OCT imaging was performed. **Fig. 6g** shows the BS-FF-OCT spectrum extracted from the green area shown in **Fig. 6c$_2$**. The spectrum shown in **Fig. 6g** is smoothed to reduce the noise level. The raw spectrum is shown in the supporting information **Fig. S2**. The peak positions (1550 $cm^{-1}$, 1640 $cm^{-1}$, 1730 $cm^{-1}$) shown in the spectrum are consistent with the peak positions for amide II (1550 $cm^{-1}$), amide I (1650 $cm^{-1}$), and the C=O band (1740 $cm^{-1}$) in protein and lipids, respectively. The other peak shown at 1460 $cm^{-1}$ is altered from the amide II band with the deuterium-oxide-based environment [52] (i.e., the water-based environment is not used due to the MIR absorption of water). The spectrum is consistent with the result in the literature [53] except for the peak at 1460 $cm^{-1}$.

## 3. Conclusion

We present a 3D chemical imaging technology termed bond-selective full-field optical coherence tomography (BS-FF-OCT). The capability of BS-FF-OCT is demonstrated on polymer samples, including 1-micron PMMA beads and polypropylene fibers, and biological samples, including mouse brain tissue, *C. elegans*, and human bladder cancer cell spheroids. Our BS-FF-OCT setup has demonstrated the ability to image bulky samples as thick as 150 μm. Furthermore, our setup is capable of imaging highly scattering samples, which is beyond the reach of phase tomography. With BS-FF-OCT, the high-density areas (cytoplasm) and the low-density areas (nucleus) inside a cell spheroid can be resolved. While the current implementation of BS-FF-OCT lacks the capability of resolving sub-cellular details, the resolution can be potentially improved by adopting the dynamic FF-OCT. [54]

To reveal more details in the cytoplasm, the resolution can be potentially improved by adopting the dynamic FF-OCT [54]. In summary, we demonstrate a bond-selective OCT technique that enables label-free volumetric spectroscopic imaging at isotropic 1-micron resolution, with potential broad applications in biological imaging.

## 4. Methods

### 4.1 BS-FF-OCT Setup

A schematic of the BS-FF-OCT setup is shown in Fig. 1a. The full-field optical coherence tomography (FF-OCT) is based on a Michelson interferometer. A broadband light-emitting diode (LED, UHP-T-545-SR, Prizmatix) provides Köhler illumination in both the sample and reference arms. Air objectives (SLMPLN50X, Olympus) are used in both arms. A CMOS camera (BFS-U3-17S7, FLIR) captures the widefield interferometric image. The MIR beam comes from a mid-infrared optical parametric oscillator (Firefly-LW, M Squared Lasers), tunable from 1320 cm$^{-1}$ to 1775 cm$^{-1}$. The laser outputs a 20 kHz MIR pulse train. Then, the 20 kHz MIR pulse train is modulated at 50 Hz by an optical chopper system (MC2000B, Thorlabs). The modulated MIR beam is focused by an off-axis parabolic mirror (MPD019-M03, Thorlabs) at the same side of the sample as the LED illuminates. The MIR pulse, LED probe pulse, optical chopper, and camera are synchronized by a pulse generator (9254-TZ50-US, Quantum composers) similar to the widefield MIP microscopy [27]. The reference mirror is installed on a piezo stage (MIPOS 100 SG RMS, Piezosystem Jena) to shift the phase difference between the two arms. Both reference mirror (with the piezo stage) and sample are installed on motorized stages (Z825B, Thorlabs) to achieve automated and synchronized coherence and focal plane matching for volumetric image acquisition.

### 4.2 Automatic multi-depth scanning

The coherence plane shifting in FF-OCT is critical to match objective focal and coherence planes. [7, 55] The coherence plane shift and its correction are shown in **Fig. S6**. When the system is imaging a specific depth of a sample, the coherence plane has to overlay with the focal plane (**Fig. S6a**). Then motor 1 scans the sample to the next depth. The coherence plane shifts and doesn't overlay with the new focal plane (**Fig. S6b**). Then, motor 2 has to scan a certain distance of the reference mirror to make the coherence plane overlay with the new focal plane (**Fig. S6c**). Software is developed to achieve automatic volumetric data acquisition in BS-FF-OCT. The software can automatically correct the coherence plane position by linearly shifting the reference mirror position at each depth during the multi-depth scanning, i.e., shifting $\Delta z/n$ at each depth in an (n+1)-depths multi-depth acquisition, where $\Delta z$ is the reference mirror shifting distance between the initial depth and the final depth. Manual correction is needed only at the initial depth and the final depth. The coherence plane can be corrected by linearly shifting the reference mirror position because the correction distance of the coherence plane has a linear relation with the sample shifting distance, as shown in the following equation, [7]

$$\Delta z_{coherence\ plane} = \Delta z_{sample} \cdot \frac{n_{sample}^2 - n_{immersion}^2}{n_{sample} \cdot n_{immersion}} \quad (1)$$

where the $n_{sample}$ is the refractive index of the sample, and $n_{immersion}$ is the refractive index of the immersion medium. $n_{immersion}$ is a constant and $n_{sample}$ can be treated approximately as a constant for a common sample that usually does not contain large refractive index changes within the data acquisition depth range.

### 4.3 Theory and image reconstruction

The theory of the image reconstruction process at a specific depth of the sample is summarized below. First, the phase difference change induced by the piezo stage is discussed in the "cold" state. Assuming the phase difference between the sample arm and the reference arm when the piezo stage is at its first position is $\varphi_{cold}$, when the piezo stage position changes $\Delta z$, the corresponding phase difference change is as follows,

$$\Delta \varphi = \frac{2\pi}{\lambda} \cdot \Delta z \quad (2)$$

where λ is the illumination wavelength. Since the experimental setup uses broadband LED as the light source, the effective value of λ in equation (2) can't be simply determined. The setting value of $\Delta z$ is experimentally calibrated to make $\Delta \varphi = \frac{\pi}{2}$.

Having this phase difference change equals to $\frac{\pi}{2}$, assuming that $E_{sample}^{cold}$ is the reflected field magnitude from a specific depth of the sample, that $I_{incoherent}^{cold}$ is the reflection intensity from the sample depths that are not coherent with the reference mirror, that $E_{reference}$ is the reflected light field from the reference mirror, and that $I_1^{cold}$ to $I_4^{cold}$ are the intensity of the four raw images captured by the camera with different phase differences, then $I_1^{cold}$ to $I_4^{cold}$ can be expressed as follows,

$$I_1^{cold} = I_{incoherent}^{cold} + E_{sample}^{cold} \cdot E_{reference} \cdot \cos(0 + \varphi_{cold}) \tag{3}$$

$$I_2^{cold} = I_{incoherent}^{cold} + E_{sample}^{cold} \cdot E_{reference} \cdot \cos(\frac{\pi}{2} + \varphi_{cold}) \tag{4}$$

$$I_3^{cold} = I_{incoherent}^{cold} + E_{sample}^{cold} \cdot E_{reference} \cdot \cos(\pi + \varphi_{cold}) \tag{5}$$

$$I_4^{cold} = I_{incoherent}^{cold} + E_{sample}^{cold} \cdot E_{reference} \cdot \cos(\frac{3\pi}{2} + \varphi_{cold}) \tag{6}$$

To retrieve $E_{sample}^{cold}$, we subtract equation (5) from (3) and subtract equation (6) from (4),

$$I_1^{cold} - I_3^{cold} = E_{sample}^{cold} \cdot E_{reference} \cdot 2 \cdot \cos(\varphi_{cold}) \tag{7}$$

$$I_2^{cold} - I_4^{cold} = E_{sample}^{cold} \cdot E_{reference} \cdot 2 \cdot [-\sin(\varphi_{cold})] \tag{8}$$

In equations (7) and (8), the incoherent intensity term that is from other depths is canceled but the phase term still exists. To cancel the phase term, the square of equations (7) and (8) are summed as follows:

$$\left(I_1^{cold} - I_3^{cold}\right)^2 + \left(I_2^{cold} - I_4^{cold}\right)^2 = E_{sample}^{cold} \cdot E_{reference} \cdot 4 \tag{9}$$

Since the reflected field from the reference mirror is uniform and can be treated as constant, $E_{sample}^{cold}$ can be obtained from equation (9),

It is noteworthy that only the reflected field magnitude from a specific depth, $E_{sample}^{cold}$, is used to yield the final photothermal signal, and the phase term, $\varphi_{cold}$, is canceled in equation (9). Thus, the photothermal effect from other depths, which makes the accumulated phase in the "hot" state become $\varphi_{hot}$, instead of $\varphi_{cold}$, does not contribute to the photothermal OCT signal.

Similarly, in the "hot" state, the reflected field magnitude $E_{sample}^{hot}$ can be obtained by the following equation,

$$\left(I_1^{hot} - I_3^{hot}\right)^2 + \left(I_2^{hot} - I_4^{hot}\right)^2 = E_{sample}^{hot} \cdot E_{reference} \cdot 4 \tag{10}$$

Subtract equation (10) from (9), and the depth-resolved photothermal image from the sample's depth i can be obtained,

$$(E_{sample}^{cold} - E_{sample}^{hot}) \cdot E_{reference} \cdot 4 =$$

$$\left(I_1^{cold} - I_3^{cold}\right)^2 + \left(I_2^{cold} - I_4^{cold}\right)^2 - \left(I_1^{hot} - I_3^{hot}\right)^2 - \left(I_2^{hot} - I_4^{hot}\right)^2 \tag{11}$$

Equation (11) describes how a photothermal image of a specific depth of sample can be obtained by using 4 cold raw images and 4 hot raw images.

### 4.4 Sample preparation

Polymethyl methacrylate (PMMA) beads embedded in agar gel sample preparation process is as follows. 1 mg agarose powder (Ultrapure Agarose, 16500-500) is measured and blended with 800 μL DI water and 200 μL 1 μm PMMA bead suspension (Phosphorex, MMA1000). Then the suspension is heated on a 95 °C hot plate until the agarose powder is melted. One 50 μm thick spacer is put on top of a $CaF_2$ substrate. Then the $CaF_2$ substrate with the space and a

CaF2 coverslip are preheated to 95 °C to avoid instant solidification when the hot agar gel suspension contacts with the cold $CaF_2$ substrate or coverslip. The temperature of the sample suspension and the $CaF_2$ substrate has to be below 100 °C to avoid water boiling during sample preparation. 50 μL hot sample suspension is dropped on the $CaF_2$ substrate, and then the $CaF_2$ coverslip is put on top of the $CaF_2$ substrate to sandwich the sample suspension. Finally, the sample cools down at room temperature and solidifies.

The polypropylene fiber mattress sample is made by peeling off the melt-blown fabric layer from a regular surgical mask. Then the polypropylene fiber layer is fixed on a silicon substrate by double-sided tape.

The mouse brain tissue, *C. elegans*, and T24 human bladder cancer cell spheroids sample are prepared as follows. First, the fresh mouse brain (Charles River Labs Inc, BIOSPECIMEN - BRAIN - MOUSE) is fixed in 10% formalin and sliced into 150-μm-thick slices. The wild type *C. elegans* adults and T24 human bladder cancer cell spheroids are fixed in 10% formalin. Then the samples are washed in $D_2O$-based phosphate-buffered saline (PBS) buffer three times. Then, the washed samples are sandwiched between the $CaF_2$ substrate and the $CaF_2$ coverslip. Finally, the gap between the substrate and the coverslip is sealed with nail polish.

### 4.5 Images denoising

The BM4D denoising method is by an open-source demo software for BM4D volumetric data denoising (release ver. 3.2, 30 March 2015). [56] The parameter values used are as follows. Noise standard deviation given as the percentage of the maximum intensity of the signal, 11%; noise distribution is Gaussian; BM4D parameter profile, modified profile; enable Wiener filtering; verbose mode; enable sigma estimation.

### 4.6 FTIR measurement

The FTIR spectrum is measured by a commercial FTIR spectroscopy (Nicolet FT-IR with ATR), which is a high-end optical benchtop system with 0.09 $cm^{-1}$ resolution and continuous dynamic alignment. This unit allows AutoTune and automated continuously variable aperture adjustment. A horizontal attenuated total reflectance (HATR) accessory is also available.

### 4.7 Spectrum smoothing

The Gaussian-weighted moving average filter used in this work is realized by the "smoothdata" function in MATLAB R2021b. "Gaussian" window is chosen.

## 5. Back matter

### 5.1 Funding

This work is supported by R35GM136223 and R33CA261726 to JXC.

### 5.2 Competing interests

The authors declare no competing interests.

### 5.3 Author contributions

C.Y., M.S.Ü., and J.X.C. proposed the idea of BS-FF-OCT. M.S.Ü. and J.X.C. supervised the research team. J.X.C. and M.S.Ü. revised the final version of the manuscript. C.Y. designed and built the visible probing part of the experiment setup, wrote the single-depth cold FF-OCT image acquisition function of the data acquisition software, and performed initial cold FF-OCT imaging experiments. H.Z. designed and built the MIR part of the experiment setup, wrote the photothermal image acquisition, multi-depth scanning, and hyperspectral scanning functions of the data acquisition software, and performed BS-FF-OCT imaging experiment on polypropylene fiber. H.Z. and J.Z. optimized the MIR beam optical path and performed BS-